# X-RAY QUASI-PERIODIC OSCILLATIONS AND ESTIMATES OF THE MASS AND THE SPIN OF THE NEUTRON STAR IN 4U 1728-34

**Ivan Zhivkov Stefanov[1]**

[1] Department of Applied Physics, Faculty of Applied Mathematics and Informatics,
Technical University of Sofia,
8, Snt. Kliment Ohridski Blvd.,1000 Sofia, Bulgaria
e-mail: izhivkov@tu-sofia.bg:

*Abstract: Three of the averaged light curves from the X-ray spectrum of 4U 1728-34 contain triads of quasi-periodic oscillations – two kHz, an upper and a lower one, and one low-frequency. In the present work, with the assumption that the triads occur simultaneously and with the application of the relativistic precession model the mass and the spin of the neutron star in the atoll source 4U 1728-34 is obtained. The consistency of the three independent estimates of the mass and the spin is also tested. According to the obtained results, the studied neutron star has relatively high mass and moderate spin.*

## 1. Introduction

According to the modern understanding, the quasi-periodic oscillations (QPOs) present in the power density spectra of many low-mass X-ray binaries (LMXBs) are manifestations of physical processes from the immediate vicinity of the compact objects located in them – neutron stars or black holes. Hence, QPOs provide invaluable information about strong field gravity and its sources. Moreover, it appears that QPOs, especially kHz QPOs, are one of the few currently available methods for the measurement of the spin of compact objects.

Motta et al. [1] demonstrated that the presence of a simultaneous triad of QPOs in the power density spectrum of a black hole, consisting of one low-frequency quasi-periodic oscillation and a pair of kHz QPOs – a lower and an upper, allow us to obtain very precise estimates on the mass and the spin of the central black hole. They applied the relativistic precession model (RP) [2, 3] and assumed Kerr metric for the description of the spacetime of the black hole in the microquasar GRO J1655-40. In order to evaluate the uncertainties on the estimated parameters, mass and spin, they used a Monte Carlo technique.

In a later work, using the same model and metric, Bambi [4] reanalyzed the simultaneous triad of QPOs displayed by GRO J1655-40 and proposed an alternative, numerically much more efficient method for the obtaining of estimates and confidence limits on the black hole parameters.

In both of these studies an assessment of the adequacy of the model, in the sense of goodness of fit test, is missing.

The current work aims at an application of the RP model for the estimation of the mass and the spin of the neutron star in the LMXB 4U 1728-34 through the method proposed by Bambi [4]. As a byproduct, with this object we have the possibility to test the adequacy of the model (or the metric). The power density spectrum of 4U 1728-34 contains not one but three supposedly simultaneous triads of QPOs. Each one of them can be used for the estimation of the parameters of the neutron star, so we have three independent measurements. A natural question in this situation is whether these three independent measurements comply with each-other. A possible conflict between them could be attributed to a wrong choice of either a model, or a metric, or both.

## 2. Observational data

The X-ray spectrum of the atoll source 4U 1728-34 was analyzed in [5] and [6]. The light curve which presents the X-ray variability of 4U 1728-34 is divided into 256-s long segments, observations. The power density spectra of these segments are grouped in 19 intervals according to the state of the source which is determined by its position on the color-color diagram. Each interval contains 10 to 134 spectra, which are averaged at the end. According to [5], the change of the selection criteria, which determine the inclusion of a given observation in one or another interval, does not significantly change the averaged light curves.

Three of the 19 intervals contain triads of QPOs consisting of one low-frequency quasi-periodic oscillation and a pair of twin kHz QPOs. They are reproduced here in Table 1.

**Table 1** *Observed values of the QPOs of 4U 1728-34, given in reference [6]*

| interaval | LF, (Hz) | L, (Hz) | U, (Hz) |
|---|---|---|---|
| 10 | $42.14 \pm 0.77$ | $513 \pm 18$ | $849.5 \pm 2.0$ |
| 11 | $45.52 \pm 0.66$ | $561 \pm 11$ | $875.7 \pm 1.6$ |
| 12 | $46.70 \pm 0.91$ | $604 \pm 14$ | $907.6 \pm 2.5$ |

## 3. Relativistic precession model

According to the RP model the X-ray emission is modulated by the motion of hot inhomogeneities dubbed hot spots in the accretion disk. The two kHz QPOs, the upper and the lower, and the low frequency QPOs are manifestations of, respectively, the orbital $\nu_\phi$, the periastron $\nu_\phi - \nu_r$ and the nodal precession $\nu_\phi - \nu_\theta$ frequencies of the hot spot. The epicyclic frequencies for evaluated with Kerr metric are [7, 8]

$$v_\phi = \left(\frac{c^3}{2\pi GM}\right)\frac{1}{r^{3/2} \pm a}, \tag{1}$$

$$v_r^2 = v_\phi^2 \left(1 - \frac{6}{r} - \frac{3a^2}{r^2} \pm \frac{8a}{r^{3/2}}\right), \tag{2}$$

$$v_\theta^2 = v_\phi^2 \left(1 + \frac{3a^2}{r^2} \mp \frac{4a}{r^{3/2}}\right). \tag{3}$$

The upper (lower) sign corresponds to prograde (retrograde) direction of rotation of the hot spot. In this paper all the masses are scaled with the Solar mass $M_\odot$, the radii are scaled with the gravitational radius $r_g \equiv GM/c^2$, and the specific angular momentum $a \equiv cJ/GM^2$, where $G$ is Newton's gravitational constant, $c$ is the speed of light, is the angular momentum of the star, is used.

### 4. Method

Bambi [4] defines the following merit function

$$\chi^2(a,M,r) = \frac{\left(v_{LF}(a,M,r) - v_{LF}^{obs}\right)^2}{\sigma_{LF}^2} + \frac{\left(v_L(a,M,r) - v_L^{obs}\right)^2}{\sigma_L^2} + \frac{\left(v_U(a,M,r) - v_U^{obs}\right)^2}{\sigma_U^2} \tag{4}$$

In order to obtain the best estimates of the parameters one usually minimizes the merit function with respect to them. In this case, however, the minimized merit function is not a chi-square variable since it is a function of three parameters and has exactly three terms. If it was a chi-square variable it would have zero degrees of freedom which is meaningless. The value of the minimized $\chi^2$ cannot be used for the evaluation of the goodness of the fit. It tells us nothing about the adequacy of the model. Besides, its value is zero whenever the algebraic system defined by the three observed frequencies has a solution, as mentioned in [4]. Nevertheless, we can use the merit function (4) to find the optimal values (point estimates) of the parameters of the model $a$, $M$ and $r$, i.e. the values which minimize the merit function $\chi^2$.

In order to obtain a confidence level region in the plane determined by two of the parameters, e.g. $a$ and $M$, we shift the parameters from their optimal values and minimize the quantity $\Delta\chi^2 = \chi^2 - \chi^2_{\min}$, where $\chi^2_{\min}$ is the minimum value of $\chi^2$, with respect to the third parameter, which in this case is $r$ [9]. The minimization of $\Delta\chi^2$ with respect to one of the parameters is a constraint which reduces the number of degrees of freedom by one. The presence of three data points (three QPOs in each of the observational intervals) and one constraint means that in this case $\Delta\chi^2$ has two degrees of freedom $dof = 2$ and the values of $\Delta\chi^2$ which correspond to confidence levels 68.3%, 95.4% and 99.73% are, respectively, 2.3, 6.17 and 11.8 [9].

### 5. Estimations of the mass and the spin of 4U 1728-34

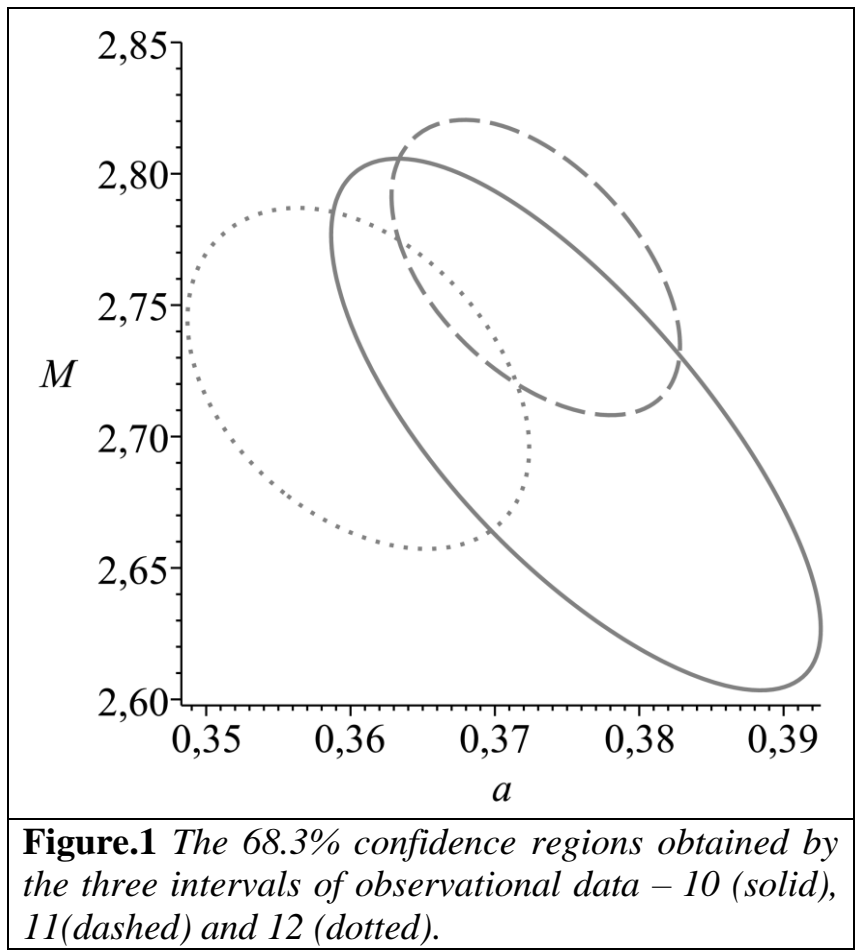

**Figure.1** *The 68.3% confidence regions obtained by the three intervals of observational data – 10 (solid), 11(dashed) and 12 (dotted).*

The 68.3% 2D confidence regions that we obtain[1] are given on Figure 1. The contours which represent the confidence regions coming from the three intervals of observational data – 10, 11 and 12 are, respectively, solid, dashed and dotted[2]. The optimal values of the fitted parameters $a$ and $M$, and their standard errors, obtained for $\Delta\chi^2 = 1$ are given in Table 2.

**Table 2** *Estimates on the mass and the spin of the neutron star in 4U 1728-34*

| interval | $a$ | $M$ |
|---|---|---|
| 10 | $0.375^{+0.012}_{-0.011}$ | $2.70 \pm 0.07$ |
| 11 | $0.373^{+0.007}_{-0.006}$ | $2.77 \pm 0.04$ |
| 12 | $0.360^{+0.008}_{-0.008}$ | $2.72 \pm 0.04$ |

### 6. Consistency test

If we treat the different observation intervals as separate, independent measurements one might want to know whether the estimates coming from them are consistent with each other. Can they be represented by a single number such as their weighted mean? In order to answer this question we define the following chi-square variables[3]

$$\chi^2_M = \sum_{i=1}^{3}\left(\frac{\bar{M}_w - M_i}{\sigma_{M,i}}\right)^2, \quad (1)$$

$$\chi^2_a = \sum_{i=1}^{3}\left(\frac{\bar{a}_w - a_i}{\sigma_{a,i}}\right)^2, \quad (2)$$

$$\chi^2_{Ma} = \chi^2_M + \chi^2_a, \quad (3)$$

---

[1] For an explanation on the obtaining of confidence limits on the fitted parameters in the case of nonlinear models we refer the reader to Section 15.6 of [9] but also Chapter 11.5 of [10]. For other pedagogical texts see also [11] and [12].

[2] The boundaries of the confidence level regions in the $a-M$ plane are implicitly defined by the system $\Delta\chi^2 = 2.3$, $\Delta\chi^2_{,r} = 0$. (Here the comma designates the derivative with respect to the parameter $r$.) This system has also geometric interpretation. It gives the projection of the 3D region defined by equation $\Delta\chi^2 = 2.3$ on the $a-M$ plane.

[3] A wonderful example of the application of the chi-square test to assess the consistency of different measurements can be found in [11].

where $\bar{M}_w$ and $\bar{a}_w$ are the weighted mean values of the mass and spin, $M_i$ and $a_i$ denote the optimal values of mass and spin coming from the $i$-th observation group, and $\sigma_{M,i}$ and $\sigma_{M,i}$ are their uncertainties. See Table 2.

We use the first two of them to assess the consistency of, the three mass estimates and the three spin estimates, respectively. The third one can be applied to test the hypothesis that both the masses and the spin are compatible with their weighted means. The values of $\chi^2_M, \chi^2_a$ and $\chi^2_{Ma}$ that we obtain, and the corresponding $p$-values evaluated with them are

$$\chi^2_M = 0.865, \quad \text{dof} = 2, \quad p-\text{value} = 0.649$$
$$\chi^2_a = 1.832, \quad \text{dof} = 2, \quad p-\text{value} = 0.400$$
$$\chi^2_{Ma} = 2.697, \quad \text{dof} = 4, \quad p-\text{value} = 0.610$$

There is no reason to question the agreement between the three independent measurements since all the $p$-values are significantly greater than 0.05.

## 7. Conclusion

According to the obtained results, the studied neutron star has relatively high mass and moderate spin. As it can be seen, the three confidence regions partially overlap, i.e. we have no obvious reason to question the applicability of the relativistic precession model to the simultaneous triads of QPOs.

Had the obtained masses been lower than $2M_\odot$, the validity of the Kerr metric for the description of the space-time in the vicinity of the neutron star in 4U 1728-34 would have been questionable.